\documentclass[journal,twoside,web]{ieeecolor}
\usepackage{jsen}
\usepackage{cite}
\usepackage{multirow}
\usepackage[T1]{fontenc}
\usepackage{amsmath,amssymb,amsfonts}
\usepackage{algorithmic}
\usepackage{graphicx}
\usepackage{hyperref}
\usepackage{textcomp}
\usepackage{bm}
\usepackage{mathtools, fixmath}
\usepackage{algorithm}
\usepackage{algorithmic}
\newcommand{\norm}[1]{\left\lVert#1\right\rVert}
\DeclareMathOperator*{\argminB}{argmin} 
\usepackage{wrapfig}
\def\BibTeX{{\rm B\kern-.05em{\sc i\kern-.025em b}\kern-.08em
    T\kern-.1667em\lower.7ex\hbox{E}\kern-.125emX}}
    
\definecolor{abstractbg}{rgb}{0.89804,0.94510,0.83137}
\begin{document}
\title{Data-Driven Nonlinear TDOA for Accurate Source Localization in Complex Signal Dynamics}
\author{Chinmay~Sahu,
        Mahesh~Banavar,~\IEEEmembership{Senior Member,~IEEE,}
        and~Jie~Sun
\thanks{Submitted on 4th August 2023 }
\thanks{C. Sahu and M. Banavar are with the Department
of Electrical and Computer Engineering, Clarkson University, Potsdam,
NY, 13676 USA e-mail: \{sahuc, mbanavar\}@clarkson.edu. J. Sun is with  Research Institute of Intelligent Complex Systems, Fudan University, Shanghai 200433, China.  e-mail: riosun@gmail.com.}}

\IEEEtitleabstractindextext{%
\fcolorbox{abstractbg}{abstractbg}{%
\begin{minipage}{\textwidth}%
\begin{wrapfigure}[13]{r}{3.5in}%
\includegraphics[width=3.1in]{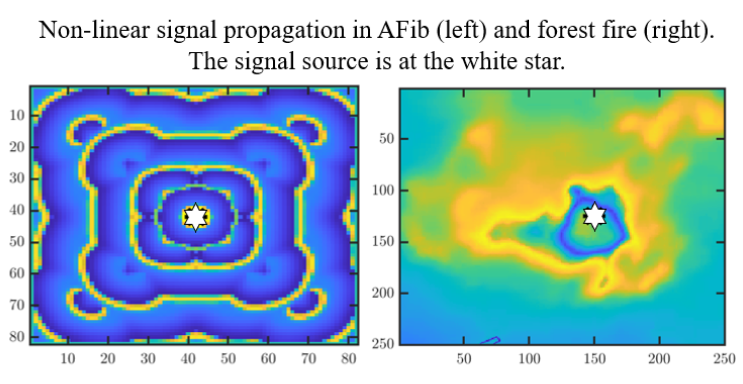}%
\end{wrapfigure}%
\begin{abstract}
The complex and dynamic propagation of oscillations and waves is often triggered by sources at unknown locations. Accurate source localization enables the elimination of the rotor core in atrial fibrillation (AFib) as an effective treatment for such severe cardiac disorder; it also finds potential use in locating the spreading source in natural disasters such as forest fires and tsunamis. However, existing approaches such as time of arrival (TOA) and time difference of arrival (TDOA) do not yield accurate localization results since they tacitly assume a constant signal propagation speed whereas realistic propagation is often non-static and heterogeneous. In this paper, we develop a nonlinear TDOA (NTDOA) approach that utilizes observational data from various positions to jointly learn the propagation speed at different angles and distances as well as the location of the source itself. Through examples of simulating the complex dynamics of electrical signals along the surface of the heart and satellite imagery from forest fires and tsunamis, we show that with a small handful of measurements, NTDOA, as a data-driven approach, can successfully locate the spreading source, leading also to better forecasting of the speed and direction of subsequent propagation.
\end{abstract}

\end{minipage}}}

\maketitle

\section{Introduction}
\label{sec:introduction}
{\IEEEPARstart{T}{arget} location estimation from data (source and target localization) using radio and acoustic signals has been subject to research for decades and it continues to receive interest in the signal processing community in research related to radar, sonar, multimedia, underwater acoustic communication, animal tracking, mobile communications, wireless sensor network, and GPS \cite{pivato2011accuracy,hasan2017survey,santos2019novel}. Still, complexities in signal propagation through non-homogeneous media \cite{poursheikhali2015tdoa,viani2011localization} together with noise and uncertainty in received signal strength makes the quality of target localization a challenging task in many complex and dynamical systems.

Localization is crucial in most wireless sensor network applications, with common methods including time of arrival (TOA), time difference of arrival (TDOA), angle of arrival (AOA), and received signal strength (RSS). Hybrid algorithms, such as large aperture arrays, have been explored to enhance signal measurement quality \cite{Manikas2012}. Among these methods, TOA and TDOA are the most versatile and widely used.

TOA typically estimates distance by using one-way ranging and the signal propagation speed from the transmitter. When the signal transmission start time is known in a synchronous network, TOA implementation is straightforward. However, most networks are asynchronous, meaning the signal transmission start time is usually unknown, rendering TOA ineffective in such cases \cite{vaghefi2013asynchronous}. Here, TDOA becomes relevant, utilizing the differences in signal arrival times at various anchor nodes for target localization.

Previous works on TDOA-based localization have often employed iterative algorithms like Maximum Likelihood Estimation (MLE) and constrained optimization techniques \cite{lin2013new, vankayalapati2014tdoa}. Later, simpler approaches, such as matrix inversion, have also gained recognition \cite{torrieri1984statistical, ho1993solution}. Extensions combining TDOA with advanced estimation methods, such as adaptive robust particle filtering (ARPF) \cite{huang2023robust} and Fixed Point Iteration (FPI) \cite{zou2023source}, and other localization techniques, including AOA, RSS measurements, and ultra-wide-band (UWB) protocols \cite{zou2023source, song2023combined, huang2023robust}, have been introduced to enable localization in non-line-of-sight (NLOS) scenarios.

However, in this paper, we develop localization methods that can handle unknown and varying propagation speeds in non-homogeneous environments, which traditional TDOA cannot solve. TDOA (multilateration) is a well-established localization algorithm known for its simplicity and effectiveness across various applications \cite{wang2019robust, li2022multi}. It estimates the time differences of a line-of-sight (LOS) signal's arrival between a reference anchor node and other anchor nodes, assuming a constant signal propagation speed, to locate the signal source \cite{tepedelenlioglu2012performance, song2023combined}.

Some existing research aims to improve the precision of TDOA estimation (see for example, \cite{yang2023polynomial}). However, these methods are confined to scenarios characterized by a constant propagation speed. In recent studies \cite{zou2020tdoa,qi2020semidefinite}, novel approaches for TDOA localization under unknown, and constant, signal speeds were presented. 

The challenge of target localization becomes apparent in cases of unknown and varying signal propagation speeds, particularly in dynamic media, where propagation speeds differ across locations due to non-homogeneous propagation media. For example,  in atrial fibrillation (AFib), where surgical ablation therapy aims to eliminate affected cardiac tissue, electrical signals propagate across the surface of the heart at unknown and varying speeds. In these cases, identifying signal sources is complex due to the intricate heart structure and spiral wave patterns \cite{jalife2003rotors}. An example AFib signal simulated with the FitzHugh-Nagumo model is shown in Figure \ref{fig:CardiacCellSimulation} \cite{you2017demonstration}. In \cite{you2017demonstration}, a triangulation-based localization algorithm is proposed, which is a variant of the time difference of arrival (TDOA) algorithm. Here, three anchors (i.e. sensors) are placed at known locations in the path of the spiral waves. The speed of propagation of the waves is assumed to be known. The arrival times of the waves at the probes are recorded, and using the speed of the propagating wave and the locations of the probes, the center of the spiral wave is estimated as the intersection of two hyperbolas associated with two pairs of probes. It is important to note that the authors in \cite{you2017demonstration} continue to assume constant propagation speed, similar to the work in \cite{zou2020tdoa,qi2020semidefinite}.

\begin{figure}[ht!]
\centering
\includegraphics[width=0.45\textwidth]{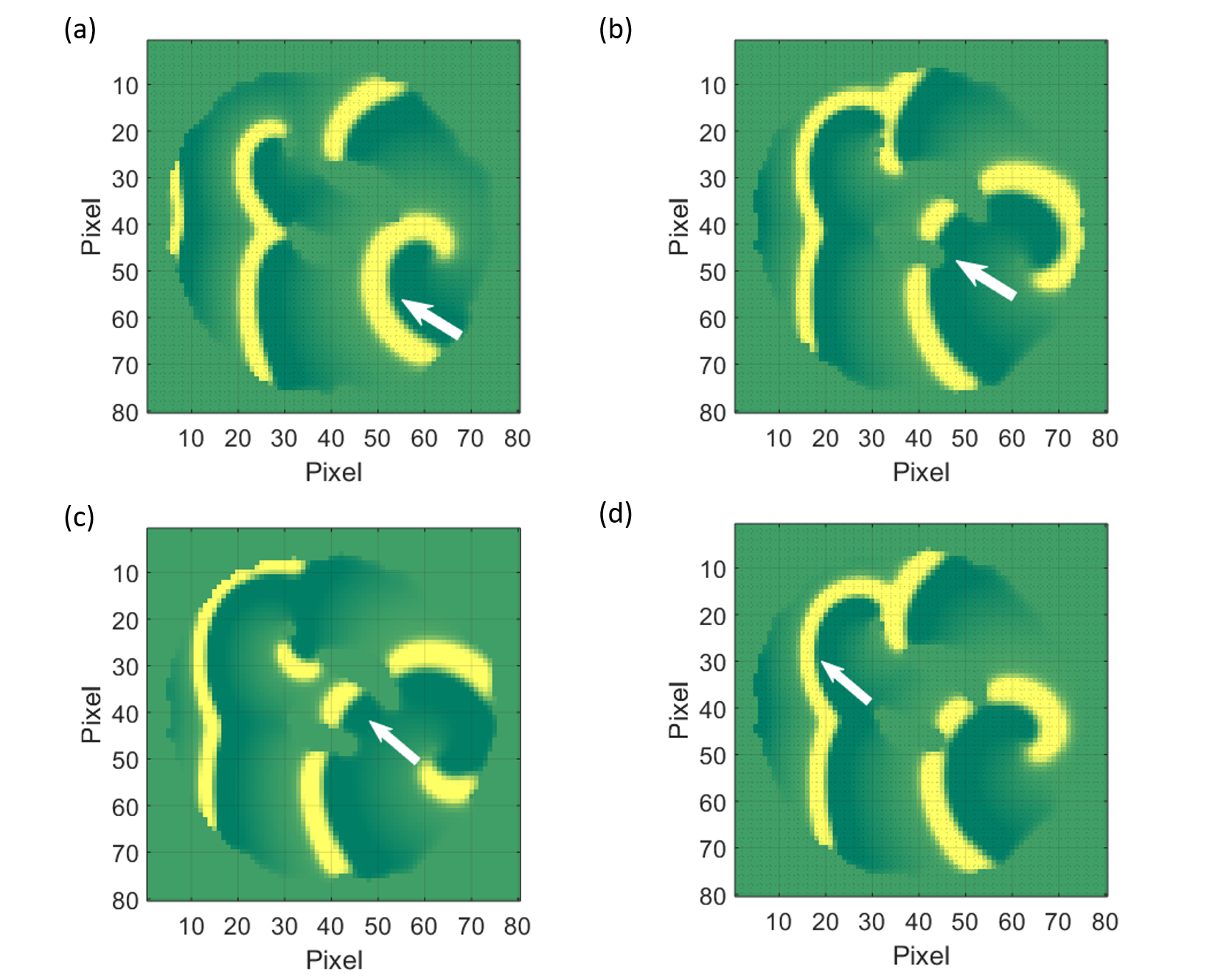}
 \caption{A cardiac wave is simulated using a modified FitzHugh–Nagumo model \cite{you2017demonstration}. Each sub-figure shows an evolution of the system, with the rotor model being visible in the stripe identified by the white arrow. }
 \label{fig:CardiacCellSimulation}
\end{figure} 

Non-homogenous media exist not just on the surface of the heart, but also in application areas such as the spread of forest fires, which can vary based on wind direction, wooded areas, and terrain. Remote sensing via computers and satellites is limited due to smoke occlusion and low camera resolution, hindering wildfire suppression. Unmanned aerial vehicles (UAVs) equipped with computer vision and GPS systems are used to detect and manage wildfires, which is a complex task as forests are non-structured environments \cite{yuan2015survey}.  Having vision sensors and GPS systems to determine fire origin location is intricate and very few researchers are working on this problem \cite{torresan2017forestry}.

Tsunami localization is a crucial research area, where wave velocity can vary due to underwater topography changes \cite{kuzakov2019localization}. Tsunameters are used for early detection and real-time measurements of deep-ocean tsunamis to improve our understanding of these phenomena and develop effective mitigation strategies \cite{gonzalez2005nthmp}. Though there are numerous methods \cite{tsushima2014tfish,tanioka2018near,voronina2016new,ariyoshi2014detectability,won2011three} to estimate source location and wave velocity, they are inherently expensive and complex in implementation \cite{tuna2017survey}.

In summary, while TDOA does not require knowledge of the signal transmission start time for source localization, it cannot accurately localize an unknown target without knowledge of the speed of the signal. It also fails to solve for an unknown target in a non-homogeneous medium. This necessitates the development of localization methods that can handle unknown and varying propagation speeds in non-homogeneous environments. 

In this paper, we present algorithms modified TDOA (mTDOA) and non-linear TDOA (NTDOA) that jointly estimate the signal source, the origin time, and the speed of propagation. Our non-linear TDOA algorithm (NTDOA) generalizes TDOA, allowing it to solve localization problems without prior knowledge of the signal origin time or the varying propagation speeds in non-homogeneous media.

Our contributions are: 
\begin{itemize}
   \item We introduce a novel modification to the traditional TDOA algorithm, enabling the solution of source localization problems where the propagation speed is unknown.
    \item We propose non-linear TDOA (NTDOA), to tackle complex source localization problems in non-homogeneous and dynamic environments.
    \item Our algorithms can be extended to estimate additional parameters, including the propagation speed. We demonstrate the effectiveness of our proposed algorithms with three real-life complex dynamical problems: atrial fibrillation, forest fires, and tsunamis.
\end{itemize}
 
The rest of this paper is organized as follows. We provide 
the problem statement in Section \ref{sec:problemStatement}. The proposed solutions are in Section \ref{sec:soltuions} and are validated numerically in Section \ref{sec:results}. Finally, concluding remarks and future work are presented in Section \ref{sec:conclu}.

\section{Problem Statement}
\label{sec:problemStatement}
The models described for phenomena such as tsunamis, forest fires, and electrical flow in AFib, are typically modeled using complex dynamical systems described by coupled differential equations. In order to derive our estimators, we use an observation-based model across all these types of systems, irrespective of the actual generation model. In order to do this, we define the forward models that the observations most closely resemble, leading to the derivation of the estimator as the inverse problem. In what follows, we describe these forward models, starting from a simple isotropic model with known propagation speed to an isotropic model with a fixed, but unknown speed, and finally, a model where the speed of propagation varies at different points in the medium.

In this article, the following notations are introduced and will be used consistently throughout:
\begin{itemize}
\item unknown source, $\mathbold{r}_0\in\mathbb{R}^3$
\item anchors at known locations, $\mathbold{r}_l$ with $l=1,\dots,N$
\item (observed) signal arrival time, $t_l$ ($l=1,\dots,N$) which represents the time at which each anchor detects the arrival of the signal from the propagation source
\item (unobserved) anchor-to-source distances, denoted as $d_l=\|\mathbold{r}_l-\mathbold{r}_0\|$ (unknown)
\end{itemize}

The general problem can be described as follows. Given the location of $N$ anchors denoted as denoted as $\{\bm{r}_l\}_{l=1}^{N}$, with observed signal arrival time $\{t_l\}_{l=1}^{N}$. The goal is to estimate the unknown location of the propagation source $\mathbold{r}_0$.

\subsection{Case-I (Isotropic medium with known speed of propagation)}
\label{ssec:isotropic model}

In the ideal case, we assume that a radiating source at an unknown location $\mathbold{r}_0\in\mathbb{R}^3$  is propagating a signal in all directions in an isotropic medium, with the signal propagation speed $c$ known and constant (see Figure \ref{fig:forward_model}(a)). Signal transmission starts at an unknown time $t_0$. There are $N$ anchors placed at known locations. The $l$-th anchor, located at $\mathbold{r}_{l}$ detects the signal at time $t_l=t_0 + d_l/c$, where  $d_l= \norm{\mathbold{r}_l-\mathbold{r}_0}$. 
In most cases we encounter, the measurements of $\{t_{l}\}$ are noisy. With these noisy $t_{l}$ values and the locations of the anchors known, our task is to estimate the location of the radiating source. Note that in this case the problem can be solved by the classical time delay of arrival algorithm \cite{sahu2019modified,tepedelenlioglu2012performance}.

\begin{figure*}[htb!]
\centering
\includegraphics[width=0.65\textwidth]{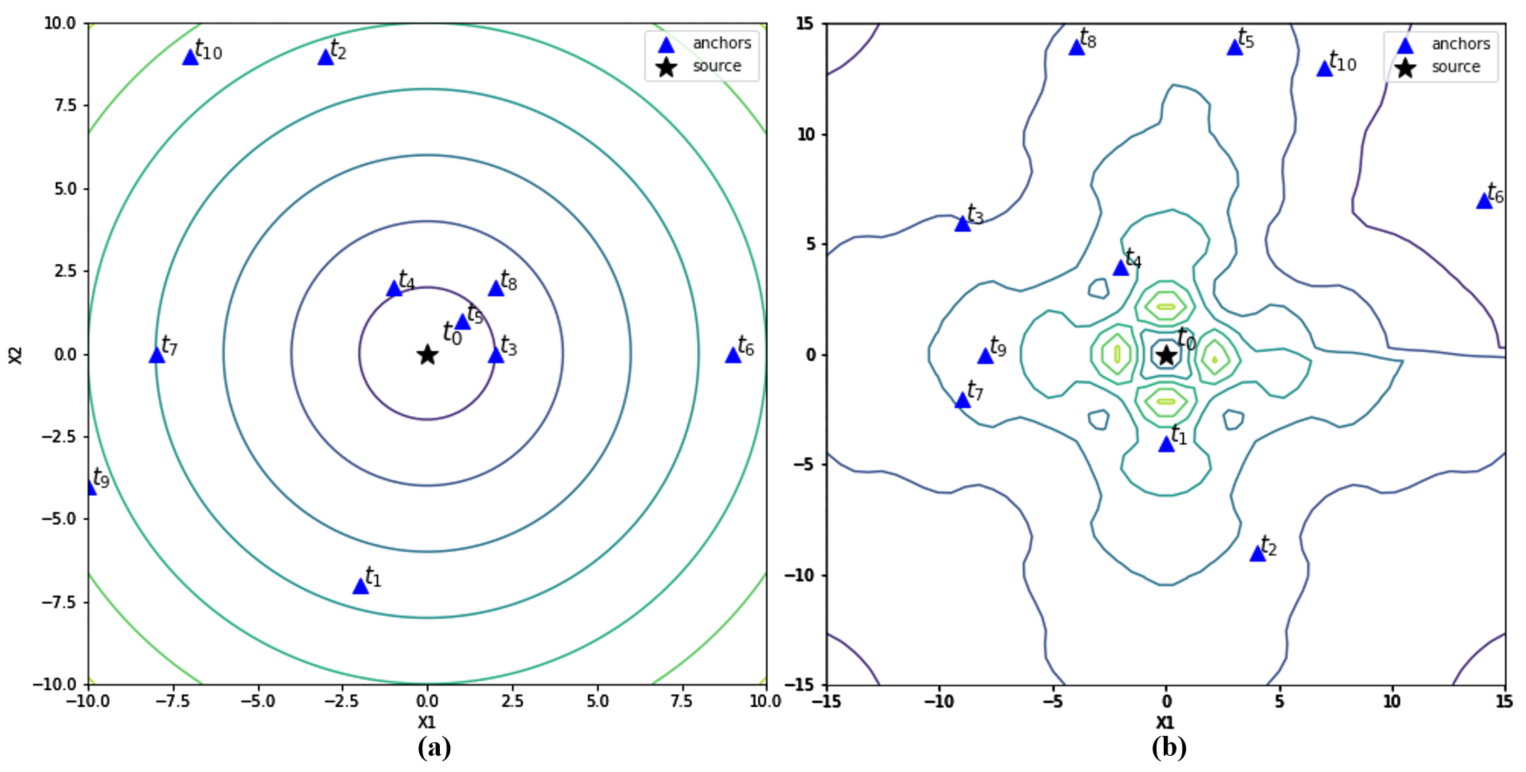}
 \caption{(a) The source (star in the middle) starts transmitting at an unknown time, $t_{0}$. The speed of propagation of the signal is fixed at all points. For Case-I (Section \ref{ssec:isotropic model}), it is assumed that the speed of propagation is known, and for Case-II (Section \ref{ssec:mTDOA}), it is unknown.  The signal is received at the anchors at times $t_{l}$, with $t_{3}$ = $t_{4}$, since they are at the same distance from the source. In the inverse problem, with the anchor locations and times $t_{l}, l \neq 0$ known, we estimate the location of the source, and the time $t_{0}$, when the source begins transmission. Additionally, for Case-II, the speed of propagation is also estimated. (b) The source (star in the middle) starts transmitting at time $t_{0}$. The speed of propagation of the signal can vary across the medium. The signal is received at the anchors at times $t_{l}$. In this case, it is not required that $t_{1}$ = $t_{4}$, even though they are at the same distance from the source since the speed of propagation can vary between the source and each anchor. In the inverse problem, with the anchor locations and times $t_{l}, l \neq 0$ known, we estimate the source location, the speed of propagation at each point, and $t_{0}$.}
 \label{fig:forward_model}
\end{figure*}

\subsection{Case-II (Isotropic medium with unknown, but constant speed of propagation)}
\label{ssec:mTDOA}

In the second, more advanced case, we drop the assumption that the speed of propagation is known while maintaining the assumption that the speed is constant. 
Here, the radiating source located at $\mathbold{r}_0$ is propagating a signal in all directions in an isotropic medium, with an unknown, but constant signal propagation speed $c$ (see Figure \ref{fig:forward_model}(a)). Signal transmission starts at an unknown time, $t_0$. There are $N$ anchors placed at known locations. The $l$-th anchor, located at $\mathbold{r}_{l}$, detects the signal at time $t_l=t_0 + d_l/c$, where  $d_l= \norm{\mathbold{r}_l-\mathbold{r}_0}$. 

\subsection{Case-III (Anisotropic medium with unknown speed of propagation)}
\label{ssec:gTDOA}
In the third and most challenging case, we consider the most practical setup by assuming that the signal propagation speed is not known and is variable due to the presence of non-homogeneous media in the transmission field. 

Here, a source  at $\mathbold{r}_0$  starts transmitting at time $t_{0}$. There are $N$ anchors placed at known locations. The speed of propagation of the signal can vary across the medium (see Figure \ref{fig:forward_model}(b)) due to it being non-homogeneous. The $l$-th anchor, located at $\mathbold{r}_{l}$ detects the signal at time:
\begin{equation}
    t_l=t_0 + t_{0,l} (\mathbold{r}_0,\mathbold{r}_l,\mathbold{s}_{0,l}),
\end{equation}
where  $t_{0,l}(.)$ is a non-linear function that calculates the propagation time between the source $\mathbold{r}_0$, the anchor $\mathbold{r}_l$, and the path between them, $\mathbold{s}_{0,l}$.

\begin{algorithm}[H]
\caption{NTDOA Algorithm}
\begin{algorithmic}[1]
    \STATE \textbf{Input:}
    \begin{itemize}
        \item Anchor positions: $\mathbold{r}_l$ for $l = 1, 2, \ldots, N$
        \item Signal reception times: $t_l$ for $l = 1, 2, \ldots, N$
    \end{itemize}
    \STATE \textbf{Output:} 
    Estimated source at $\mathbold{r}_0$, Signal  speed $c(\mathbold{r}_l,\theta)$
    \FOR{each sensor $l$ }
        \STATE Calculate Signal reception  time $ t_l $
    \ENDFOR
    \STATE Estimate $\mathbold{r}_0$ using Eqn. (\ref{eqn:NTDOA})
    \STATE Solve the system of equations using a numerical solver
    \STATE \textbf{Return} estimated source $\mathbold{r}_0$,  Signal speed  $c(\mathbold{r}_l,\theta)$
\end{algorithmic}
\label{algo:NTDOA}
\end{algorithm}

\section{Solutions}
\label{sec:soltuions}
In the earlier section, we described forward models, starting from a simple isotropic model with known propagation speed, to an isotropic model with a fixed, but unknown speed, and finally, a non-isotropic model, where the speed of propagation can vary at different points in the medium. In what follows, we propose solutions for each case leading to the derivation of the estimator as the inverse problem together with computational methods of solving them.

\subsection{Case-I (Isotropic medium with known speed of propagation)}
\label{ssec:isotropic_solutions}

\subsubsection*{Inverse Problem}
With $N$ anchors, the unknown target location and unknown start time can be estimated by solving:
\begin{equation}
\mathbold{x}= \argminB_{\mathbold{r}_{0}, t_{0}} \sum_{l=1}^{N} \left[ c^2 (t_l-t_o)^2-\norm{\mathbold{r}_{l}-\mathbold{r}_{0}}^2\right]^2,
\label{eqn:TDOA}
\end{equation}
where $\mathbold{x}$ contains the estimates of the target location $\mathbold{r}_{0}$, and the unknown start time $t_{0}$. The location of the $l$-th anchor is $\mathbold{r}_l$, and the time at which the signal passes it is $t_{l}$. 

Note that this optimization problem also has a matrix solution given by  \cite{sahu2019modified,tepedelenlioglu2012performance}:

\begin{equation}
   \mathbold{x}=\mathbold{H}^{\#}\mathbold{b}. 
    \label{eqn:leastSquareSol}
\end{equation}
where the vector of unknowns $\mathbold{x}$ contains the target location $\mathbold{r}_{0}$ and the unknown start time $t_{0}$. $\mathbold{H}^{\#}$ is the pseudo-inverse of $\mathbold{H}$, which contains time difference data and the corresponding distances between anchors. $\mathbold{b}$ is the measurement vector from $N$ anchors containing anchor location information. Since this is the classical TDOA problem, the solution is well known \cite{sahu2019modified,sahu2020localization}.   The solution discussed here using Eqn. (\ref{eqn:leastSquareSol}) is not necessarily optimal when the observations are noisy \cite{UNLOC2018}.

\subsection{Case-II (Isotropic medium with unknown constant speed of propagation)}
\label{ssec:mTDOA_algo}

\textit{Inverse problem:}  With $N$ anchors placed in the anchor field around the source,  the proposed  modified TDOA (mTDOA) algorithm jointly estimates the source location, speed of propagation, and initial time of the signal by solving the following optimization problem:  
\begin{equation}
\mathbold{x}= \argminB_{\mathbold{r}_{0}, t_{0}, c} \sum_{l=1}^{N} \left[ c^2 (t_l-t_o)^2-\norm{\mathbold{r}_{l}-\mathbold{r}_{0}}^2\right]^2,
\label{eqn:mTDOA-NM}
\end{equation}
where $\mathbold{x}$ contains the estimates of the target location $\mathbold{r}_{0}$, unknown signal propagation speed $c$, and unknown start time $t_{0}$. The location of the $l$-th anchor is $\mathbold{r}_l$, and the time at which the signal passes it is $t_{l}$. 
The optimization problem in \eqref{eqn:mTDOA-NM} can be solved using numerical methods such as the Nelder-Mead \cite{singer2009nelder} algorithm, or in matrix form as in Eqn. (\ref{eqn:leastSquareSol}), with the modification that the vector of unknowns $\mathbold{x}$, contains the target location $\mathbold{r}_{0}$, the unknown start time $t_{0}$, and the unknown signal propagation speed $c$ \cite{sahu2019modified,sahu2020localization}.

\subsection{Case-III (Anisotropic medium with unknown speed of propagation)}
\label{ssec:NTDOA_Solution}
\textit{Inverse problem:} With the anchor locations and times $t_{l}, l \neq 0$ known, we estimate the location of the source, the speed of propagation at each point of interest, and the value of $t_{0}$, when the source begins transmission.
The earlier inverse solution considered an unknown and constant signal propagation speed during problem formulation, which fails to work for a signal having variable propagation speed. Therefore, we reformulate the optimization problem stated in Eqn. (\ref{eqn:mTDOA-NM}) to 
 \begin{equation}
\bm{x}^{\text{(NTDOA)}}= \argminB_{r_{0}, t_{0}, c} \sum_{l=1}^{N} \left[ c(\mathbold{r}_l,\theta)^2 (t_l-t_o)^2-\norm{\mathbold{r}_{l}-\mathbold{r}_{0}}^2\right]^2,
\label{Eqn:Goptimization}
\end{equation}

The speed at which the measured signal propagates, $c(\mathbold{r}_l,\theta)$, varies across different anchor locations $\mathbold{r}_{l}$ due to the presence of non-homogeneous media and is dependent on the direction. It is not a fixed value.  Unlike standard TDOA where speed is constant (linear), in this more general framework Eqn. (\ref{Eqn:Goptimization}) models it as a {\it nonlinear} function that varies in spatial coordinates, leading to the nonlinear TDOA (NTDOA) algorithm.

While there are several approaches that can be used to model the non-linear function in Eqn. (\ref{Eqn:Goptimization}), here, we make the simplifying assumption that we decompose the non-linearity into the polar form, and represent $c(\mathbold{r}_l,\theta)$ as a product:
\begin{equation}
    c(\mathbold{r}_l,\theta) \approx f(R_l)g(\theta_l),
\end{equation}
where $(R_l=\|\bm{r}_l-\bm{r}_0\|,\theta_l)$ represents the polar coordinates of $\bm{r}_l-\bm{r}_0$, and $f$ and $g$ are nonlinear functions. Here, $f$ and $g$ can be modeled per the non-homogeneous propagation medium. In an ideal isotropic medium, $g(\cdot) = 1 $ and $f(\cdot) = c$, and, Eqn. (\ref{eqn:NTDOA}) reduces to Eqn. (\ref{eqn:mTDOA-NM}).

One simple approach is to model it using the Taylor series and the Fourier series, respectively. Here, $f$ represents non-linearity in speed as a function of radius which we further approximate using Taylor series as $f(R) \approx \sum_{k=0}^{K} a_k R^k$; $g$ encodes speed non-linearity as a function of angle, which we represent using Fourier series as
$g(\theta) \approx 1+ \sum_{\ell=1}^{L} b_\ell \cos(\omega_\ell \theta)+ d_\ell \sin(\omega_\ell \theta)$. Together, $f$ and $g$ encompass a large class of nonlinear models to capture the non-homogeneity of propagation speed in space. The resulting optimization problem now reads
\begin{equation}
\begin{aligned}
& \bm{x}^{\text{(NTDOA)}} =\argminB_{ t_0, \bm{r}_0\in\mathbb{R}^2, \{a_k\}_{k=0}^{K}, \{\omega_\ell, b_\ell, d_\ell\}_{\ell=0}^{L} }  \sum_{i=1}^{N} \\
&
\Bigg[ \left(\sum_{k=0}^{K} a_k R_i^k\right)^2  \\
&
\left( 1 + \sum_{\ell=1}^{L} b_\ell \cos(\omega_\ell \theta_i)+ d_\ell \sin(\omega_\ell \theta_i)\right)^2 (t_i-  t_0)^2.\\
&-\
 \left.\|\bm{r}_i - \bm{r}_0\|^2 \right.\Bigg]^2,
\end{aligned}
\label{eqn:NTDOA}
\end{equation}
where $\bm{r}_i = (R_i\cos(\theta_i), R_i\sin(\theta_i))\in\mathbb{R}^2$. The total number of unknowns is $K+3L+4$. Here, for simplicity, we assume the lowest-order nonlinear model using $K=L=1$, with 8 unknowns and at least 9 anchors required to solve the problem. As the complexity of the medium changes, the values of $K$ and $L$ can be adjusted to derive better model fits.

In the simplest case, the NTDOA algorithm involves the determination of 8 unknown variables in a non-linear optimization problem. When using numerical optimization methods to solve for these variables, multiple local minima may be obtained, or in some cases, the numerical algorithms may not converge. The initialization of the numerical algorithms is, therefore, important. In order to converge to the global solution, which is a solution that is valid for the entire set of data and not just a subset, we initialize the numerical algorithm being used to solve the NTDOA algorithm with the estimates from the mTDOA algorithm. This ensures that the first guess for the NTDOA algorithm is close enough to the true solution that we have convergence and to the global solution. 

\section{Results}
\label{sec:results}

The proposed modified time difference of arrival (mTDOA) and non-linear TDOA (NTDOA) algorithms can be used to solve localization problems where the start time, signal propagation speed, and target location are unknown. These algorithms have several potential applications, including the estimation of the spiral wave core in atrial fibrillation, the determination of the origin and speed of propagation of a forest fire, and the estimation of the source and speed of propagation of a tsunami. In each of these cases, the algorithms can be used to accurately and efficiently estimate the relevant parameters, enabling a better understanding of the underlying phenomena and potentially enabling more effective response or prevention efforts. In what follows, we discuss the application of the proposed algorithms to atrial fibrillation, wildfires, and tsunamis. 

\subsection{Numerical Models and Data}
\label{ssec:ModelSimnData}
To demonstrate the effectiveness of the proposed algorithm, we made use of three different datasets. One is the simulation data obtained from the FitzHugh-Nagumo (FHN) model, where the FHN model captures the dynamics of the heart. The FHN model provides valuable insights into the behavior of spiral waves in the heart and can be used to explore the potential impacts of various interventions or conditions on these dynamics. The other two datasets are real-time satellite imagery from the Creek Forest Fire in California in 2020 and the Tonga tsunami in 2022.

We used a modified version of the FitzHugh-Nagumo (FHN) model \cite{fitzhugh1961impulses,nagumo1962active} to simulate the dynamics of spiral waves in the heart. This model allows for the generation of pulsating waves that move outward from a rotor core. In our simulations, following \cite{fitzhugh1961impulses,nagumo1962active}, we used a fixed rotor core within an $80$mm$ \times 80$mm square. The evolution of model simulation is shown in Figure \ref{fig:process_evolution} (left).

\begin{figure*}[h!tb]
\centering
\includegraphics[width=0.6\textwidth]{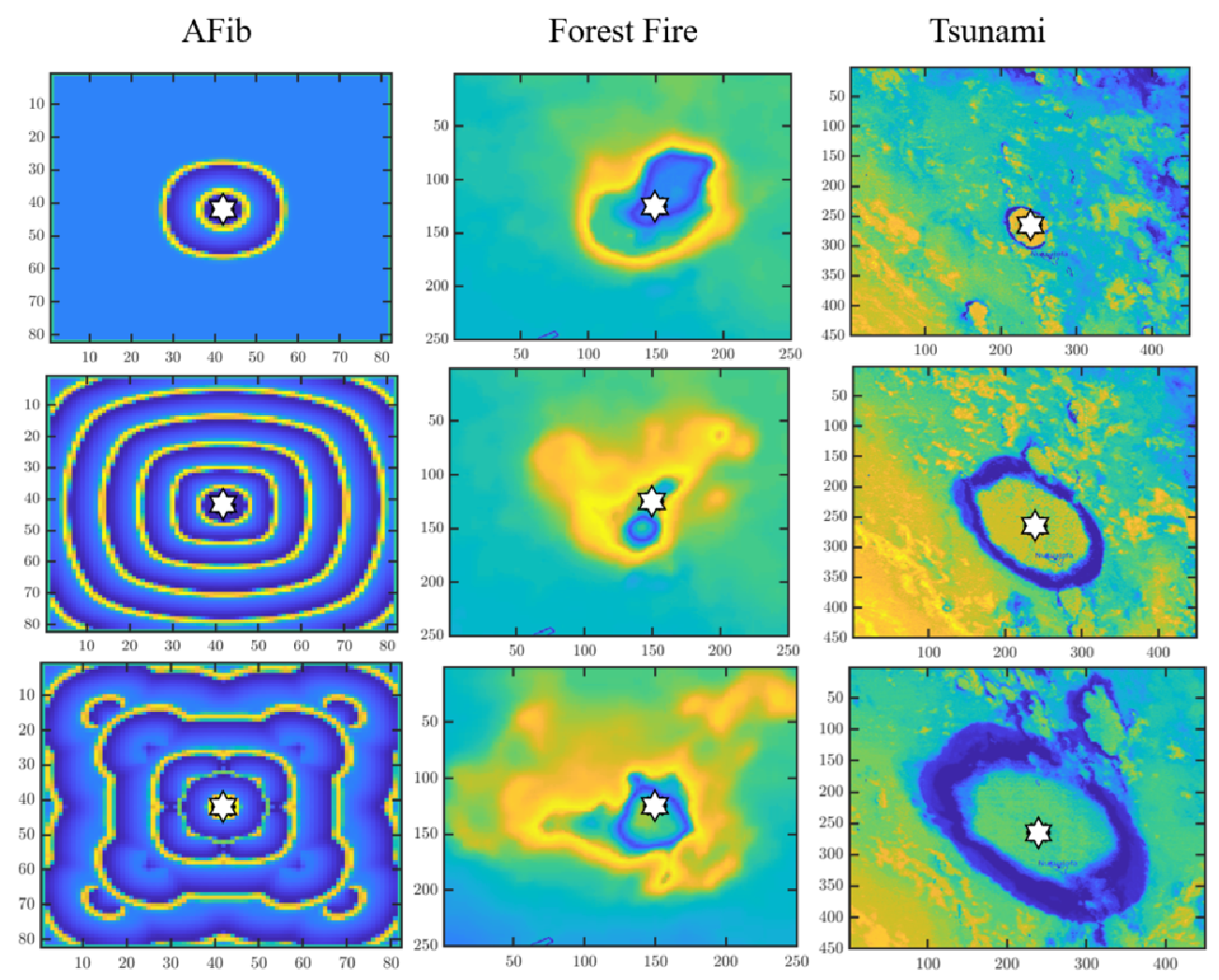}
 \caption{Signal propagation patterns in realistic scenarios. Each sub-figure (top to bottom) shows an evolution of the three dynamic processes. (L-R) AFib (modified FHN), Creek wildfire \cite{NASA, safford20222020}, Tonga tsunami \cite{omira2022global}. The Source of propagation is marked with a white hexagram in each figure.}
 \label{fig:process_evolution}
\end{figure*}

We obtained satellite imagery from NASA's Earth Observing System Data and Information System (EOSDIS) Worldview application \cite{NASA}, which was collected every 10 minutes using Band 13 during the Creek Forest Fire in California in 2020. Each pixel in the imagery represents a 2-kilometer distance, and the estimated average fire propagation speed was 6-14 miles per hour \cite{safford20222020}. A snapshot of the evolution of the forest is shown in  Figure \ref{fig:process_evolution} (middle).

Similarly, satellite imagery was collected during the 2022 Tonga volcanic eruption and tsunami. Each pixel in the image represents a 2-kilometer distance. The estimated wave propagation speed during the tsunami was 1000 kilometers per hour \cite{omira2022global}. The evolution of the wave is shown in Figure \ref{fig:process_evolution} (right).

\begin{figure*}[h!tb]
\centering
\includegraphics[width=0.8\textwidth]{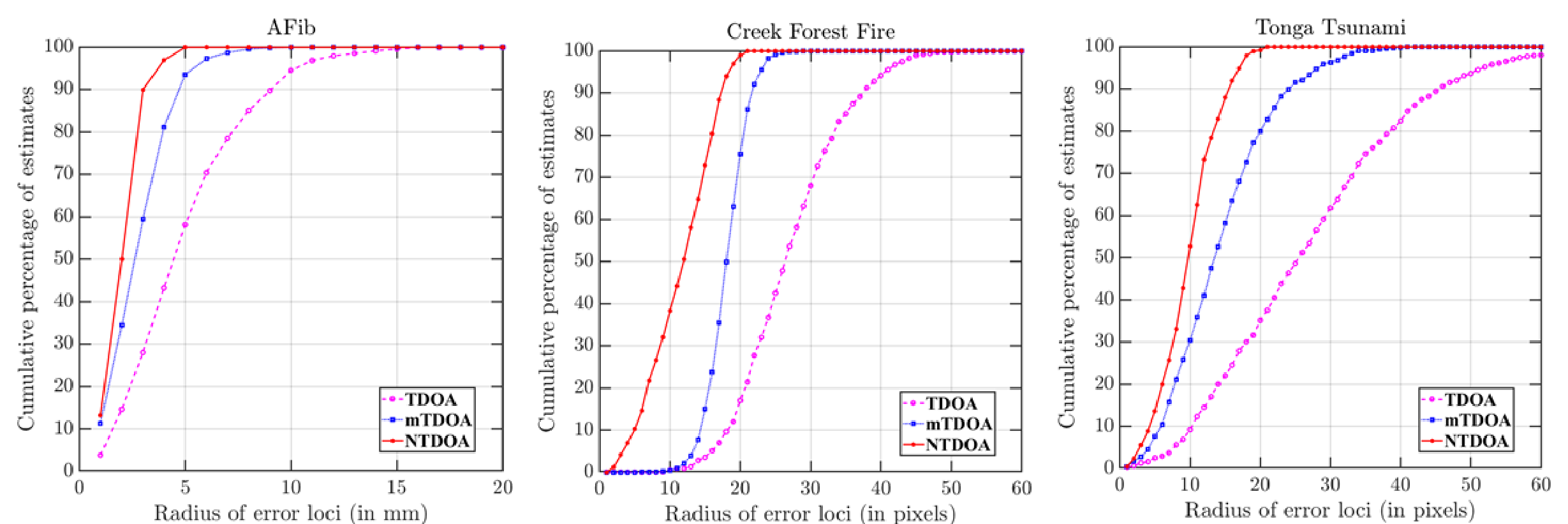}
 \caption{Localization Results using NTDOA: (Error Quantification) To estimate the accuracy of the algorithm, the number of estimates within a circle of a given radius are counted. The percentage of estimates for each radius value is plotted. The faster the curve reaches 100\%, indicates a more accurate the estimator. The algorithm is tested with 50 randomly placed anchors, repeated 1000 times, and the results are averaged. The results show that NTDOA performs better compared to other methods.}
 \label{fig:NTDOA_errQ}
\end{figure*}

\subsection{Simulation Results}
\label{results}

To assess the efficacy of the proposed algorithms, we conducted two experiments.  In the first experiment, we evaluated the performance of the TDOA, modified TDOA (mTDOA) and non-linear TDOA (NTDOA) algorithms in determining the target's location in different scenarios as discussed in Section \ref{ssec:ModelSimnData}. We calculated the discrepancy between the estimated and actual target locations and repeated this process 1000 times using 50 randomly placed anchors in each iteration. The effectiveness of the algorithms was determined by counting the number of estimates that fell within a specified radius, as shown in Figure \ref{fig:NTDOA_errQ}.

In the second experiment, the results of which are shown in  Figure \ref{fig:NTDOA_errCurve}, we varied the number of anchors from 10 to 50 in increments of 5 and estimated the mean absolute error for each evaluated method through a Monte-Carlo simulation of 1000 trials. We evaluated, in each case, the mean absolute error in the localization estimation of the signal sources. 

Finally, we estimated the average signal propagation speed for all the methods over a Monte-Carlo run of 1000 and tabulated the results in Table \ref{table:speed}. It is interesting to note that the difference in average estimated speed can be attributed to the variation in measured propagation speed at different anchor locations in each method.
\begin{figure*}[h!tb]
\centering
\includegraphics[width=0.8\textwidth]{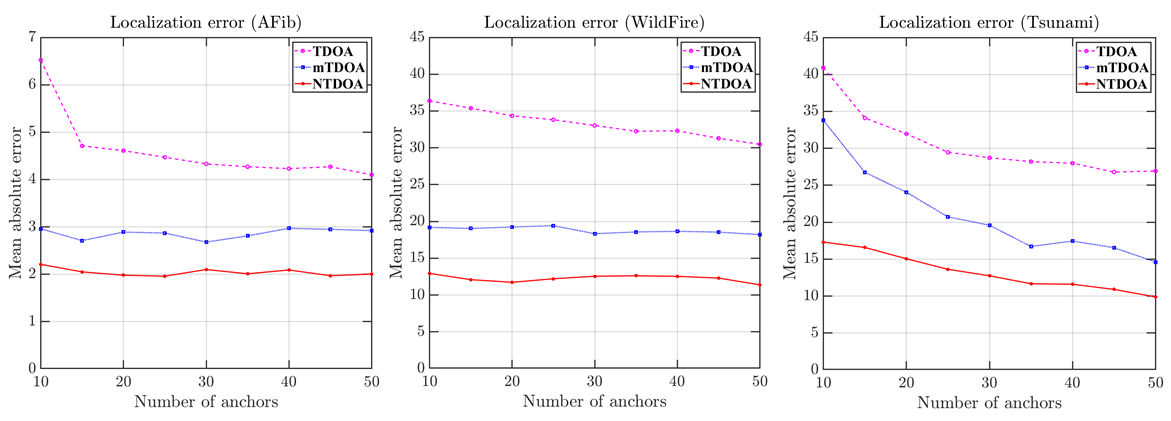}
 \caption{Localization error results using NTDOA: (Error Quantification) To quantify  the accuracy of the algorithm, we measured the mean absolute error (MAE) under varying conditions. We simulated scenarios with 10 to 50 anchors in the search space, repeating each simulation 1,000 times and averaging the results. Our findings demonstrate that NTDOA consistently achieved the lowest MAE compared to other methods.}
 \label{fig:NTDOA_errCurve}
\end{figure*}

\subsection{Discussion}
\label{ssec:discussion}

In what follows, the results presented in Section \ref{results} are interpreted and discussed. 

\subsubsection{AFib}
\label{ssec:AFib}

To localize the rotor core leading to atrial fibrillation, we utilized a fixed rotor core within an 80mm x 80mm square and positioned anchors at known locations selected at random to determine the rotor center's position. The results shown in Figure \ref{fig:NTDOA_errQ} (left) indicate that using TDOA, 100\% of the estimates were within a 15mm radius of the rotor center. The performance improved for mTDOA and NTDOA, with nearly 100\% of the estimates being within a 5mm radius for NTDOA. In Figure \ref{fig:NTDOA_errCurve} (left), the mean absolute error (MAE) was highest for TDOA, which assumed a constant speed of propagation. The MAE decreased by 50\% when the signal propagation speed was considered unknown for mTDOA, and was the lowest for NTDOA, which jointly estimates the speed and direction of propagation and the target location.

It is important to note that we used an average signal propagation speed of 0.7 m/s from literature \cite{young2010anisotropy} as the assumed propagation speed when using the TDOA algorithm. We can see from Table \ref{table:speed} that the speed of propagation estimates from mTDOA and NTDOA are closer to 0.36 m/s. This leads to a significant estimation error when using TDOA. However, NTDOA accurately captured the signal's propagating dynamics, estimated a more accurate target location, and performed the best among the evaluated methods.

\subsubsection{Forest Fire}
\label{res:fire}

To estimate the source and flow of the forest fire, we used satellite imagery that captured the images of the fire. In each frame, we use points on the boundary of the fire as indicators of progress and use these as anchor locations with the time of arrival given by the frame time-stamp. This simulates the placement of thermal sensors at these locations, which are activated as the fire reaches them. 

Using these data points, we calculate the error between the estimated and actual location of the wildfire's origin. The results, shown in Figure \ref{fig:NTDOA_errQ} (middle), demonstrate that TDOA resulted in 100\% of the estimates falling within a 35-pixel radius of the wildfire's origin, while mTDOA and NTDOA showed improved performance. With NTDOA, nearly 100\% of the estimates were within a 15-pixel radius. Overall, the NTDOA algorithm significantly reduced the error by 50\% or more compared to TDOA. Similarly, as seen in Figure \ref{fig:NTDOA_errCurve} (middle), the mTDOA method, which considered the speed of propagation as an unknown factor, resulted in a mean absolute error (MAE) that was 50\% lower than the MAE of TDOA which assumed a propagation speed of 10 km/h ($\approx$ 6.2 miles per hour \cite{safford20222020}). The NTDOA method, which continuously estimated the speed and direction of the signal as it calculated the target's location, had the lowest MAE than the other methods across all experiments.

\subsubsection{Tsunami}
\label{res:tsunami}

Similar to the approach in Section \ref{res:fire}, we use satellite imagery to simulate anchor locations in the experiments to evaluate the location of the tsunami's origin. The results presented in Figure \ref{fig:NTDOA_errQ} (right) indicate that TDOA had 100\% of its estimates within a 60 pixel radius from the actual origin, while mTDOA and NTDOA demonstrated improved performance. NTDOA was able to place almost 100\% of its estimates within a 20 pixel radius. The NTDOA algorithm showed a significant improvement of 50\% or more compared to TDOA. In a complementary experiment (as shown in Figure \ref{fig:NTDOA_errCurve} (right)), it was observed that the TDOA method, which assumed a constant signal propagation speed of 1000 km/h \cite{omira2022global}, had the highest MAE. However, when the speed of propagation was considered as an unknown in mTDOA, the MAE decreased by nearly 50\%. The NTDOA method, which continuously determines the speed and direction of the signal as it calculates the target's location, had the lowest MAE.

In this study, we proposed a novel approach to source localization with transmissions over non-homogeneous media leading to unknown and varying propagation speeds. Our approach involved reformulating non-linear signal dynamics as an inverse problem and developing a nonlinear decomposition of the in-homogeneous velocity field. With our new algorithms, we were able to achieve more accurate source localization when compared to TDOA. This enhanced accuracy can have significant implications in various fields, enabling faster response, predictive monitoring, and management. Across all experiments and applications, the results suggest that mTDOA and NTDOA can be valuable tools for precisely determining source locations and propagation speeds, potentially enabling more targeted and timely interventions.

\begin{table}[!ht]
\caption{Comparison of Signal Propagation Speeds. For TDOA, the actual velocity is used as an input to the algorithm. The proposed mTDOA and NTDOA techniques do not assume a speed of propagation. }

\label{table:speed}
\centering
\resizebox{1\columnwidth}{!}{\begin{tabular}{cccc}
\hline
Algorithm & AFib (m/s) & Forest Fire (km/hr) & Tsunami (km/hr) \\ \hline
True speed & 0.7 & 10 & 1000 \\ 
TDOA (Given)                       & 0.7        & 10                 & 1000           \\ 
mTDOA (Est.)                     & 0.364      & 8.43               & 1224           \\ 
NTDOA (Est.)                     & 0.359      & 8.41               & 1126           \\ \hline
\end{tabular}}
\end{table}

\section{Conclusions}
\label{sec:conclu}

In this paper, we presented two new algorithms for determining the origin and speed of propagation of a signal propagating over non-homogenous media, namely, modified time difference of arrival (TDOA) and non-linear TDOA (NTDOA). The algorithms have applications in various settings including in healthcare (atrial fibrillation) and geo-hazards (forest fires, tsunamis).  
The algorithms were validated using simulated and real-world data 
and the results indicate that the NTDOA algorithm was the most effective, particularly when compared to classical approaches such as TDOA. 

In future work, we plan to conduct detailed investigations into the nuances of each complex dynamical system, including time and spatial complexity analysis. We aim to explore the use of these algorithms in scenarios where the velocity of spiral waves is variable, obstacles block or alter the waves, or multiple sources are present. Additionally, we intend to modify the assumptions and parameters of the NTDOA algorithm to better suit various environmental conditions.

\section*{Acknowledgment}

We acknowledge the use of imagery from the NASA Worldview application (https://worldview.earthdata.nasa.gov/), part of the NASA 
(EOSDIS).

\bibliographystyle{IEEEtran}
\bibliography{jsen}

\vspace{-0.7in}
\begin{IEEEbiography}[{\includegraphics[width=1in,height=1.25in,clip,keepaspectratio]{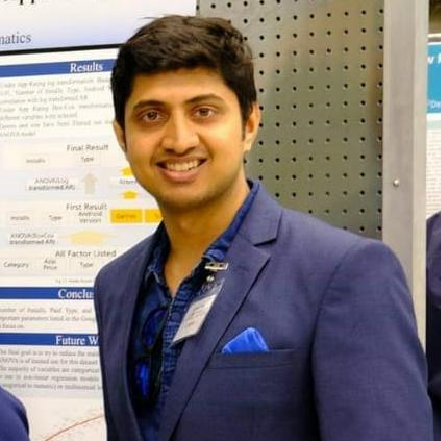}}]{Chinmay Sahu} received his B.S. from Biju Patnaik University of Technology, Odisha, India. Then he received his MS from the National Institute of Technology, Tiruchirappalli, Tamilnadu, India. Before joining his Ph.D., he worked as a Software Designer for 2 years with Alstom Transport India Ltd., Bangalore. His current research interests are in localization, machine learning, computer vision, and biometrics.
\end{IEEEbiography}

\vspace{-0.7in}
\begin{IEEEbiography}[{\includegraphics[width=1in,height=1.25in,clip,keepaspectratio]{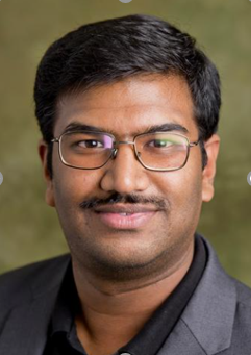}}]{Mahesh Banavar}(S'08-M'11-SM'17) received the B.E. degree in telecommunications engineering from Visvesvaraya Technological University, India, in 2005, and the M.S. and Ph.D. degrees in
electrical engineering from Arizona State University, Tempe, in 2007 and 2010, respectively. He is currently an Associate Professor with the Department of Electrical and Computer Engineering, Clarkson University, Potsdam, NY, USA. His interests include node localization, detection and
estimation algorithms, and user-behavior-based
cybersecurity applications.
\end{IEEEbiography}

\vspace{-.7in}
\begin{IEEEbiography}[{\includegraphics[width=1in,height=1.25in,clip,keepaspectratio]{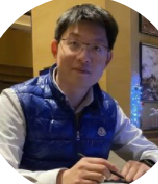}}]{Jie Sun}
 received his BS in Physics and (dual-degree) in Finance from Shanghai Jiao Tong University in 2006, and PhD in Mathematics from Clarkson University (USA) in 2009. He subsequently received postdoctoral training at Northwestern University and Princeton University, after which he joined the Department of Mathematics at Clarkson University as a faculty member in 2012. Dr. Sun joined Huawei Hong Kong Research Center as a Chief Researcher in September 2019. Having broad research interests in applied mathematics and statistical physics, drawn to the fascinating role they play in applications, he is now leading an effort to define and tackle fundamental theoretical problems in nonlinear dynamics and complex systems with applications in network optimization, data compression, and physics-inspired computation. This work was mainly done during Dr. Sun’s sabbatical leave and visit at Fudan University during 2018-2019.  
\end{IEEEbiography}

\end{document}